\documentclass[]{aa}
\usepackage{graphicx}
\usepackage{natbib}
\usepackage{color}
\bibpunct{(}{)}{;}{a}{}{,} 
\begin{document}

\def\al{$^{26}$Al}
\def\fe{$^{60}$Fe}
\def\g{$\gamma$}
\def\feal{$^{60}$Fe/$^{26}$Al}
\def\Ms{M$_{\odot}$}
\def\Zs{Z$_{\odot}$}
   \title{Radioactive \al \ and \fe \ in the Milky Way: \\
implications of the RHESSI detection of \fe }

   \author{ N. Prantzos\inst{1}
   }
   \offprints{N. Prantzos}
   \institute{Institut d'Astrophysique de Paris, 98bis Bd. Arago,
    75104 Paris, France
   }
   \date{}

   \abstract{
The recent detection of gamma-ray lines from radioactive \al \ and
\fe \ in the Milky Way by the RHESSI satellite calls for  a
reassessment of the production sites of those nuclides. The
observed gamma-ray line flux ratio is in agreement with
calculations  of nucleosynthesis in massive stars, exploding as
SNII (Woosley and Weaver 1995); in the light of those results,
this observation would suggest then that SNII are the major
sources of \al \ in the Milky Way, since no other conceivable
source produces substantial amounts of \fe. However, more recent
theoretical studies find that SNII produce much higher \feal \
ratios than previously  thought and, therefore, they cannot be the
major \al \ sources in the Galaxy (otherwise \fe \ would be detected
long ago, with a line flux similar to the one of \al). 
Wolf-Rayet stars, ejecting \al \ (but not \fe) \ in
their stellar winds, appear  then as a most natural candidate. We
point out, however, that this scenario faces also an important
difficulty. Forthcoming results of ESA's INTEGRAL satellite, as
well as consistent calculations of nucleosynthesis in massive
stars (including stars of initial masses as high as 100 \Ms \ and
metallicities up to 3 \Zs), are required to settle the issue.
\keywords{ Galaxies: Milky Way } }

   \maketitle
%

\section{Introduction}

\al \ is the first radioactive nucleus ever detected in the Galaxy
through its characteristic gamma-ray line signature, at 1.8 MeV
(Mahoney et al. 1982). Taking into account its short lifetime ($\sim$1
Myr), its detection convincingly demonstrates that nucleosynthesis
is still active in the Milky Way (Clayton 1984). The detected flux
($\sim$4 10$^{-4}$ cm$^{-2}$ s$^{-1}$) corresponds to $\sim$2 \Ms \
of \al \ currently present in the ISM (and produced per Myr,
assuming a steady state situation). The COMPTEL instrument aboard
CGRO mapped the 1.8 MeV emission in the Milky Way and found it to
be irregular, with prominent "hot-spots" probably associated
with  the spiral arms (Diehl et al. 1995). The spatial distribution of \al \
suggests that massive stars are at its origin (Prantzos 1991,
1993, Prantzos and Diehl 1996). However, it is not yet clear
whether the majority of observed \al \ originates from the winds
of the most massive stars (i.e. above 30 \Ms, evolving as
Wolf-Rayet stars) or from the explosions of less massive stars
(i.e. in the 12-30 \Ms \ range, exploding as SNII); the
uncertainties in the corresponding stellar yields are still quite
large (see Sec. 2) and do not allow to conclude yet.

 Clayton (1982)  pointed out that SNII explosions produce
another relatively short lived radioactivity, \fe \ (lifetime
$\sim$2 Myr). Since WR winds do not eject that isotope, the
detection of its characteristic gamma-ray lines 
\footnote{At 1.117 and 1.332 MeV, resulting from the decay of its daughter
nucleus $^{60}$Co}   in the
Milky Way would constitute a strong argument  for SNII being at
the origin of \al. Based on detailed nucleosynthesis calculations
of SNII (from Woosley and Weaver 1995) Timmes et al. (1995) found
that the expected gamma-ray line flux ratio of \feal \
(for each of the  two lines of \fe) is
0.16, if SNII are the only sources of \al \ in the Milky Way.

The Reuven Ramaty High Energy Solar Spectroscopic Imager (RHESSI)
detected  the galactic \al \ emission at a flux level compatible
with previous observations (Smith 2003a). Most recently, Smith
(2003b) reported the first ever detection of the Galactic \fe \
gamma-ray lines with RHESSI; their combined
fluxes correspond to a significance level slightly higher than 3
$\sigma$. The line flux ratio \feal \ is found to be 0.16 (for each
\fe \ line), precisely at the level predicted by Times et al
(1995) on the basis of Woosley and Weaver (1995) nucleosynthesis
calculations.

This finding of RHESSI appears as an impressive confirmation of a theoretical
prediction. However, more recent studies of SNII nucleosynthesis
produce different values for the \feal \ ratio (see next section),
considerably higher than the one of Timmes et al. (1995). 
Combined with the RHESSI finding, the new theoretical results
call for a reassessment of the \al \ sources in the Milky Way.
In this work we discuss those results and their implications.
We argue that none of the proposed sources of \al \ satisfies all
observational constraints at present. Forthcoming observations
by the INTEGRAL satellite, combined with a new generation of stellar
nucleosynthesis models (for rotating massive stars up to 100 \Ms \ 
and metallicities up to 3 \Zs) will probably be required to settle the issue.

\section{\al \ and \fe: revised yields and \g-ray line fluxes}

Four different groups  (to our knowledge) have performed
calculations of nucleosynthesis in massive stars, estimating the
amounts of both \al \ and \fe \ and covering a relatively extended
grid of stellar masses: Thielemann et al. (1995), Woosley and
Weaver (1995), Rauscher et al. (2002) and Limongi and Chieffi
(2003). In the first case, however, presupernova calculations are
made in pure He-cores and the amounts of hydrostatically produced
\al \ are seriously underestimated; therefore, those results are not
discissed in the following.

The calculations of Woosley and Weaver (1995, hereafter WW95) and
of Rauscher et al. (2002, hereafter RHHW02) are made with
essentially the same stellar evolution code, but the latter
benefit from improved stellar physics and, especially, an updated
library of nuclear reaction rates. Thus, the RHHW02 results
supersede those of WW95, at least for solar metallicity stars
(WW95 is the only published work  providing yields of
radioactive nuclei for an extended grid  of stellar
metallicities). Both those calculations take into account
neutrino-induced nucleosynthesis during the supernova explosion,
which increases the \al \ yield by about 40\% on average (WW95).

Finally, the Limongi and Chieffi (2003, hereafter LC03)
calculations are done with a different stellar evolution code but
with essentially the same set of nuclear reaction rates as RHHW02
(the REACLIB library of Rauscher and Thielemann). They adopt a
different treatment for the study of the explosion than RHHW02 and
they do not take into account neutrino-induced nucleosynthesis.

The situation concerning the \al \ and \fe \ yields of those
calculations is summarized in the first and second panel of Fig.
1, respectively. In the top panel, it is clearly seen that the \al
\ yields of RHHW02 are substantially smaller than those of WW95,
by a factor two on average. That difference is obviously due to
the different input physics adopted in the two studies.

The LC03 yields of \al \ are even smaller than those of RHHW02,
and that difference can be attributed, at least partially, to the
neglect of the neutrino-induced nucleosynthesis in the former
study. Note that, in order to account for the uncertainties of the
supernova explosion LC03 study a range of explosion energies, and
this affects (slightly) the \al \ yield of their lowest mass
stars. Note also the interesting "convergence"of the three
calculations in the case of the 20 \Ms \ star, perhaps because the
properties of that particular stellar mass are better constrained
after the extensive study of SN1987A.

\begin{figure}
\centering%
\includegraphics[angle=0,width=0.5\textwidth]{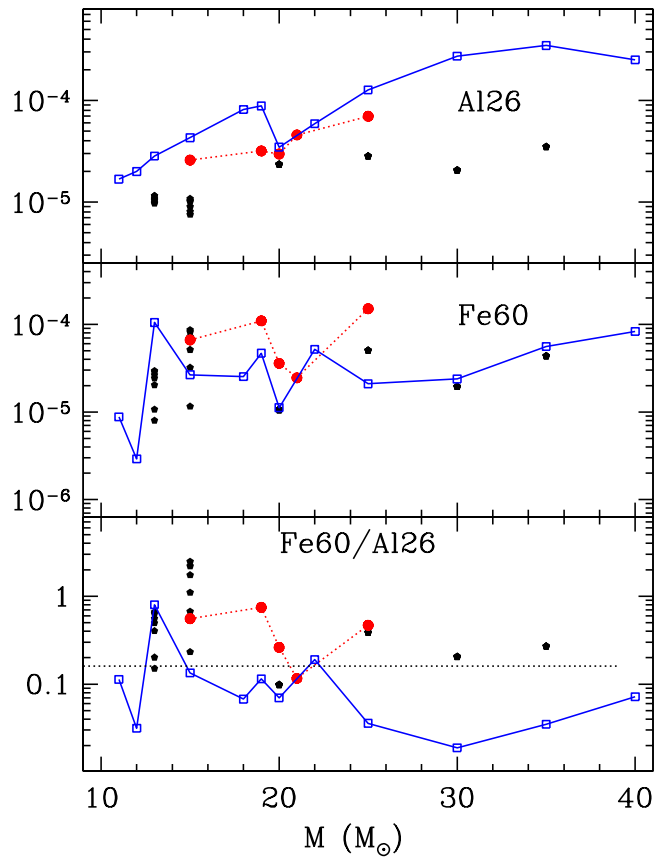}
\caption{\label{Falfe-yields} Yields (in \Ms) of \al \ ({\it top})
and \fe ({\it middle}) and the \fe/\al \ gamma-ray line flux ratio
({\it bottom}) as a function of stellar mass, according to
calculations by WW95 ({\it open squares} connected by solid
lines), RHHW02 ({\it filled circles} connected by dotted lines)
and LC03 ({\it filled pentagons}); in the latter case, different
points for a given stellar mass correspond to different explosion
energies. All calculations are for solar metallicity stars. The
{\it dotted } line in the lowest  panel corresponds to the RHESSI
value of 0.16 (Smith 2003b).}
\end{figure}

In the case of \fe, RHHW02 obtain yields twice as large as WW95,
on average. The reason for that discrepancy is probably the
improved library of nuclear reaction rates of RHHW02. Combined with
the results for \al, it becomes obvious that RHHW02 get \feal \
ratios four times larger than WW95. The corresponding results of
LC03 are in excellent agreement with WW95 above 20 \Ms \ and in
fair agreement with those of RHHW02. In the 13-15 \Ms \ range, the
\fe \ yields of LC03 depend strongly on the explosion energy, with
the lower energies leading to higher yields. Note, however, that
in all cases \fe \ is produced by successive neutron captures on
Fe-peak nuclei; the last step ($^{59}$Fe(n,$\gamma$)\fe \ involves
the unstable nucleus ($^{59}$Fe, for which there are no experimental data
concerning its neutron capture cross-section. The nuclear uncertainties
on its yield are thus quite important.\footnote{The WW95 yields
are calculated with cross section value of 1.8 mb for the
neutron capture on  $^{59}$Fe, where the RHHW02 and LC03 calculations
adopt a value of 3.4 mb, see Woosley et al. (2003).}

The corresponding ratio of \feal \ by number (i.e. the yield ratio
divided by 60/26) for each stellar mass appears in the bottom
panel of Fig. 1. The results of WW95 are, on average, close to the
value of 0.16 (dotted horizontal line),  mentioned in the RHESSI
discovery report of \fe \ (Smith 2003b), while those of RHHW02 and
LC03 are substantially above that value for almost all the stellar
masses.

To compare properly with observations, these yields should be
convolved with a stellar Initial Mass Function (IMF) and we adopt
here the Salpeter IMF, a power-law\footnote {A
power-law IMF is defined (as a function of stellar mass M) as :
$\Phi(M)=dN/dM=A M^{-(1+x)}$.} with slope x=-1.35,
in order to obtain the number ratio
\begin{equation}
R(M_{UP}) = \int_{12 M_{\odot}}^{M_{UP}} \
{{Y_{60}(M)}\over{Y_{26}(M)}} \ {{26}\over{60}} \ \Phi(M) \ dM
\end{equation}
as a function of the upper limit of integration $M_{UP}$. The
results are plotted on Fig. 2. In the case of LC03 two curves are
shown: LC03H corresponds to the high \fe \ yields (low explosion
energies) and LC03L to the low \fe \ yields (high explosion
energies). In all cases the thick portions of the curves
correspond to the stellar mass range covered by each study, while
the filled hexagons mark the highest mass of each calculation and
thus provide the IMF weighted value over the whole mass range
covered by each study.

\begin{figure}
\centering%
\includegraphics[angle=-90,width=0.5\textwidth]{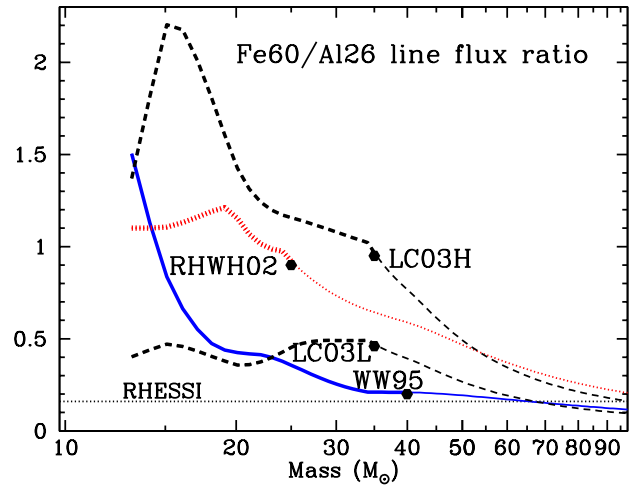}
\caption{\label{Falfe-IMF}  The expected ratio of \fe/\al \ decays
(for each of the two \fe \ lines), convolved with a Salpeter
stellar Initial Mass Function, is shown as a function of the upper
stellar mass limit of the convolution integral (see Equ. 1). The
four curves correspond to the four different sets of stellar
yields, with their thick portions corresponding to the mass range
covered in those works (see text). The {\it dotted } horizontal
line at 0.16 is the \fe/\al \ ratio reported by
RHESSI. {\it Filled pentagons} mark the upper mass in each of the
four studies; all recent calculations predict much higher values
than the older calculations of WW95 or the observed value of
RHESSI. Only by taking into account the \al \ yields of massive
Wolf-Rayet stars ({\it thin} portion of the curves beyond the masses
indicated by the filled pentagons,  obtained
by adding data for WR stars from Meynet et al.
1997) one may obtain \fe/\al \ ratios
compatible with the RHESSI results. }
\end{figure}

The WW95 yields, integrated up to the highest mass of that study
(M$_{UP}$=40 \Ms) lead to a number ratio \feal=0.18, i.e. very
close to the value 0.16 advanced by Timmes et al. (1995) on the
basis of those same yields and the same IMF. It is precisely that
theoretical prediction, well within reach of modern instruments,
that made \fe \ a prime target for astrophysical gamma-ray
spectroscopy. The RHESSI discovery apparently confirms that
prediction. However, an inspection of the more recent results
shows that the modern theoretical expectations are, in fact, much
higher: near unity for RHHW02 and LC03H and above 0.4 in the case
of LC03L. In the light of those results, the RHESSI discovery at the 
level predicted by Timmes et al. (1995) looks more as a coincidence.

Assuming that the recent theoretical results are not to be
substantially revised in the future and that the RHESSI result is
confirmed,what are the implications for our understanding of the
origin of \al? The obvious conclusion is that the bulk of galactic
\al, detected by various instruments including RHESSI, is not
produced by the source of \fe: if this were the case, then \fe \
would be detected with a line flux similar to the
one of \al. Obviously, another source
of \al \ is required, producing much smaller \feal \ ratios
than the SNII.

The obvious candidate source is Wolf-Rayet stars, as has been
argued in many places over the years (e.g. Dearborn and Blake
1985, Prantzos and Cass\'e 1986, Prantzos 1991 and  1993, Prantzos
and Diehl 1996, Meynet et al. 1997, Kn\"odlseder 1999). The winds
of those massive, mass losing stars, eject large amounts of \al \
produced through H-burning in the former convective core, {\it
before} its radioactive decay (in stars with no mass loss, those
quantities of \al \ decay inside the stellar core before the final
explosion and never get out of the star). Note that WR stars eject
negligible amounts of \fe, since that nucleus is produced at more
advanced stages of the stellar evolution than \al \ and there is
no time for it to be ejected before the final explosion (e.g.
Prantzos et al. 1987). However, no complete calculations of WR
stars (i.e. of massive stars, say above 40 \Ms, with mass loss and
up to the final explosion) are available up to now. The
calculations of Woosley et al. (1995) concern only the advanced
evolution of massive He cores and ignore any contribution of the
WR winds to the \al \ yields (besides, it is difficult to link the
mass of their calculated He cores to the mass of the corresponding
main sequence stars). Thus, the total \al \ and \fe \ yields of
those stars (i.e. the sum of the masses ejected by the winds and
by the explosion) is unknown at present. Although the number of
such stars in a normal IMF is small, the amounts of \al \ in the
winds are extremely large and affect considerably the overall
budget. In what follows, we assume that those stars eject
negligible total amounts of \fe \ (otherwise one faces the same
problem with a high \fe/\al \ ratio as before).

Adopting the \al \ yields of non-rotating WR stars of solar
initial metallicity by Meynet et al (1997), which concern stars of
solar metallicity in the 25-120 \Ms \ mass range, and combining
them with the aforementioned SNII yields, one obtains the \fe/\al
\ ratio expected by the total mass range of massive stars, during
all the stages of their evolution; this is expressed in Fig. 2 by
the continuation of the four theoretical curves above the masses
indicated by the filled pentagons. It can be seen that the RHESSI
result is recovered in that case, provided that at least half of
\al \ originates from WR stars (in the case of LC03L), or even
that 80\% of \al \ originates from WR stars (in the case of RHHW02
or LC03H).

At this point, it should be noted that the aforementioned yields
are not the most appropriate for a discussion of the galactic
\fe/\al \ ratio. Indeed,  the metallicity gradient observed in the
Milky Way disk (-0.07 dex/kpc for oxygen and several other metals,
see Hou et al. 2001 and references therein) implies an average
metallicity of around 2 \Zs \ in the present-day  disk. As already
noted in several studies (e.g. Prantzos and Cass\'e 1986), it is
the yields of stars with such a metallicity that contribute mostly
to the metal enrichment of the Milky Way today.

\begin{figure}
\centering%
\includegraphics[angle=-90,width=0.5\textwidth]{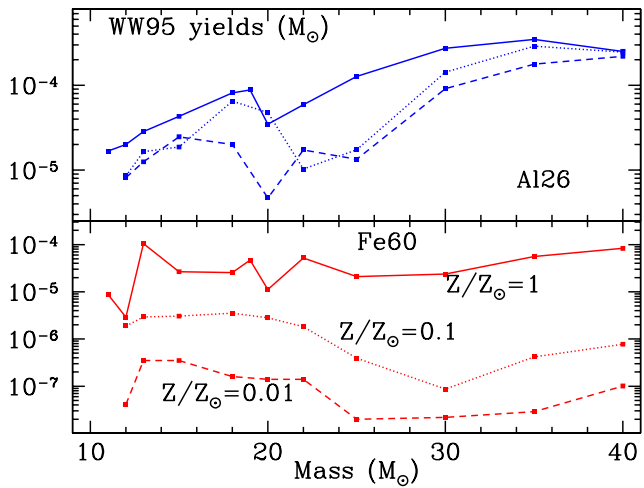}
\caption{\label{Falfe-Met}  Yields of SNII from WW95 for \al \ ({\it top})
and \fe \ ({\it bottom}), for three different values of the
initial stellar metallicity (as indicated in the bottom panel). }
\end{figure}

Unfortunately, the works of RHHW02 and LC03 cover only solar
metallicity stars, while the WW95 study considers a range of
stellar metallicities below solar. One may, however, extrapolate
from the trends obtained in the WW95 study at 2 \Zs \ and scale
accordingly the recent yields of RHHW02 and LC03. The WW95 yields
for stars of initial metallicities \Zs, 0.1 \Zs \ and 0.01 \Zs \
are displayed in Fig. 3, for \al \ (upper panel) and for \fe \
(lower panel), respectively. It can be easily see that the \fe \
yields are systematically proportional to the initial stellar
metallicity for most stellar masses; the reason is that \fe \ is
mostly produced by neutron captures in the carbon shell and its
yield is proportional to the initial $^{56}$Fe amount. On the
other hand, the yields of \al \ are slightly higher at \Zs \ than
at 0.1 \Zs. Part of \al \ is produced in the H-shell by proton
captures on initial $^{25}$Mg and this does depend on initial
metallicity; however, the bulk is produced in the C-shell
(more than 80\% in the 25 \Ms \ star; 
see, e.g. Fig. 1 in Timmes et al. 1995) , where
$^{25}$Mg is produced by $^{12}$C, itself resulting from the
initial H and He of the star, and thus it is independent of the
initial metallicity.

One concludes then that at 2 \Zs \ the \fe \ yields of SNII
(i.e. stars in the 12-25 \Ms \ range) should be on
average twice the corresponding ones at \Zs, while the \al \
yields should be only slightly higher than their counterparts at
\Zs. This implies in turn that the curves of \feal \ displayed in
Fig. 2 (corresponding to \Zs \ stars) are in fact lower limits to
the values expected from the galactic population of massive stars
\footnote {In Timmes et al. (1995) a detailed model is presented
for the Galactic disk, but the WW95 yields are used and no
explicit mention is made of the galactic metallicity gradient.
Thus, the ratio of 0.16 is obviously an underestimate, based on
yields of stars with metallicity not exceeding \Zs.}. This only
exacerbates the discrepancy between the  RHESSI
result and the theoretical expectations from SNII, and makes the
\al \ contribution of WR stars even more important. Since the \al
\ yields of WR stars increase with metallicity approximately as
Z$^{1.5}$ or Z$^{2}$(see below), they can easily match the
increased \fe \ yields of SNII at 2 \Zs \ and bring the average galactic
\feal \ ratio close to the RHESSI value. These
qualitative considerations should be substantiated, of course, by
self-consistent calculations of rotating
stars at metallicities higher than \Zs,
extended as to cover all the advanced evolutionary phases, as well
as the final explosion (see Heger et al. 2000  and Hirschi et al
2003 for  preliminary results of such calculations).

\begin{figure}
\centering%
\includegraphics[angle=-90,width=0.5\textwidth]{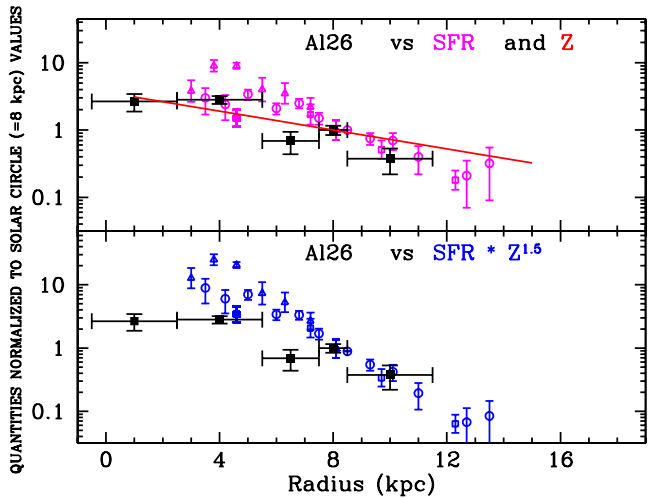}
\caption{\label{Falfe-IMF}  Radial distributions of Al26, star formation rate
(SFR) and metallicity (Z) in the Milky Way disk. {\it Upper panel:}
Data points with vertical error bars correspond to various tracers of the
SFR, while the galactic metallicity profile of oxygen (with
a gradient of dlog(O/H)=-0.07 dex/kpc) is shown by a solid line;  the
Al26 profile, after an analysis of {\it COMPTEL} \ data by Kn\"odlseder
(1997), is shown  (in relative units)
by data points with vertical and horizontal error bars (the horizontal ones
correspond to the adopted radial binning). {\it Lower panel:} If galactic
Al26 originates mostly from WR stars, its radial distribution should scale
with SFR * Z$^{1.5}$ (points with vertical error bars, scaled from the upper
panel), since the Al26 yields of WR stars scale with Z$^{1.5}$ at least
(Vuissoz et al. 2003, calculations for rotating stars);
however, the observed Al26 distribution
(same points as in upper panel) is flatter than the expected
one in that case.
}
\end{figure}

 There is another way to
understand the implications of the revised yields for the \al \
sources, which does not involve \fe. Indeed, observations of the
Galactic 1.8 MeV line by different instruments converge to a value
of 4 10$^{-4}$ photons/cm$^2$/s , which corresponds to a steady
state value of 2 \Ms \ of \al \ (produced per Myr) in the
interstellar medium (e.g. Diehl et al. 1995, Prantzos and Diehl
1996, Diehl and Timmes 1998). 
The average \al \ yield in the recent calculations of SNII
is 2.5 10$^{-5}$ \Ms (compared to  10$^{-4}$ \Ms \ in WW95).
Taking into account the average SNII frequency observed in Sbc
galaxies like our own ($\sim$1-2 per century, Cappellaro et al.
2003) one sees that SNII may produce about 0.3-0.6 \Ms \ of \al \
per Myr, i.e. about 15-30 \% of the total amount inferred from
observations. The necessity for another source of \al \ becomes
obvious again. On the basis of non-rotating stellar models, Meynet
et al. (1997) show quantitatively that WR stars can indeed provide
the bulk of galactic \al. This is also supported by a different
argument (Kn\"odlseder 1999) concerning the similarity of the
Galactic maps of \al \ and of ionizing photon flux (provided only
by the most massive stars, those that eventually become WR).
Moreover, Kn\"odlseder et al. (2001) point out that one of the
prominent "hot-spots"of the COMPTEL 1.8 MeV map, the Cygnus
region, is an association of very young massive stars, with no
sign of recent supernova activity.

Those arguments point towards WR stars as major sources of \al \
in the Milky Way. However, the situation  is far from being clear
yet, because the WR stellar yields of \al \ depend strongly on
metallicity. In the case of non rotating stellar models that
dependence is $\propto$ Z$^2$, according to Meynet et al. (1997).
The rotating models of WR stars, currently calculated by the
Geneva group (Meynet and Maeder 2003) show that rotation
considerably alleviates the need for high mass loss rates, while
at the same time leading to the production of even larger \al \
yields than the non-rotating models (Vuissoz et al. 2003); in that
case, it is found that the \al \ yields of WR have a milder
dependence on metallicity ($\propto$ Z$^{1.5}$) than the non
rotating ones. In both cases, that metallicity dependence of the
\al \ yields of WR stars, combined with the radial profiles of star
formation rate (SFR) and of metallicity in the Milky Way (see Fig.
4, upper panel) suggest that the resulting radial profile of \al \
should be much steeper than the one actually observed. The latter,
derived from COMPTEL observations (Kn\"odlseder 1997) appears in
Fig. 4 (lower panel) and is clearly flatter than the product
SFR*Z$^{1.5}$ (as already noticed in Prantzos 2002). Similar
conclusions are reached if the longitude, rather than radial,
profiles of \al, metallicity and SFR  are considered.

\section{Conclusion}

Contrary to a rather widely spread opinion, the recent RHESSI detection
of radioactive \fe \ in the Milky Way does not imply that \al \ is
mostly produced by supernova explosions. Recent theoretical results
suggest that the \fe \ line flux would
then be close to the one of \al \ (within a factor of two).
Assuming that both the RHESSI results and the recent stellar nucleosynthesis
results hold, another source of \al \ should be found.

Wolf-Rayet stars appear as natural candidates, in view of their absolute \al \
yields  (at least in the framework of the Geneva models:
either with high mass loss rates and  no rotation - Meynet et al. 1997 -
or with mild mass loss rates and rotation - Vuissoz et al. 2003)
and  presumably low \feal \ ratios.
However, the strong dependence of the \al \  yields on metallicity suggests
that the \al \ emissivity should be steeply increasing in the inner
Galaxy, while the COMPTEL observations clearly display a milder enhancement
at small Galactic longitudes.

Thus, almost twenty years after its discovery (Mahoney et al.
1982), the \al \ emission of the Milky Way has not yet found a
completely satisfactory explanation. Indeed, the recent
observational (COMPTEL, RHESSI) and theoretical (RHHW02, LC03,
Vuissoz et al. 2003) results have made the puzzle even more
complex than before. The solution will obviously require progress
in both directions. From the theory point of view, detailed
nucleosynthesis calculations of mass losing and rotating stars up
to the final explosion in the mass range 12-100 \Ms \ and for
metallicities up to 3 \Zs \ will be required ; furthermore,
the uncertainties still affecting the reaction rates of $^{22}$Ne($\alpha$,n) 
(major neutron producer
during He burning in massive stars) and $^{59}$Fe(n,$\gamma$) 
will have to be substantially reduced. From the observational
point of view, the radial distributions of both \al \ and \fe \
will be needed; such distributions will probably be available if
the operation of ESA's INTEGRAL satellite is prolonged for a few
years beyond its nominal 2-year operation.

\acknowledgements {I am grateful to Roland Diehl and to the referee,
Mark Leising, for valuable comments on the manuscript.}

\def\aj{AJ \ }
\def\apj{ApJ \ }
\def\apjs{ApJS \ }
\def\aa{A\&A \ }
\def\aas{A\&AS \ }

\bibliographystyle{apj}

\begin{thebibliography}{46}
\expandafter\ifx\csname natexlab\endcsname\relax\def\natexlab#1{#1}\fi

\bibitem[Cappelarro(2003)]{cappelarro03}
Cappellaro E., Barbon R., Turatto M., 2003, 
to appear in proceedings of IAU Colloquium 192, 
"Supernovae (10 years of 1993J)",  Marcaide J. M. and  Weiler K. W. (eds),
(astro-ph/0310859)

\bibitem[Clayton (1982)]{clayton82}
Clayton D. D., 1982, in Essays in Nuclear Astrophysics, Barnes C.
et al. (eds.), CUP, Cambridge, p. 401

\bibitem[Clayton (1984)]{clayton84}
Clayton D. D., 1984, \apj, 280, 144


\bibitem[Dearborn-blake (1985)]{dearbornblake85}
Dearborn D., Blake J., 1985, \apj 288, L21

\bibitem[Diehletal (1995)]{diehl95}
Diehl R., Dupraz C., Bennett K., et al., 1995, \aa 298, 445

\bibitem[Diehl-Timmes (1998)]{diehl98}
Diehl R., Timmes F., 1998, PASP 110, 637

\bibitem[Heger(2000)]{heger00}
Heger A., Langer N., Woosley S., 2000, \apj 528, 368

\bibitem[Hirschi(2003)]{hirschi03}
Hirschi R., Meynet G., Maeder A., Goriely S., 2003, 
to appear in proceedings of IAU Colloquium 192, 
"Supernovae (10 years of 1993J)",  Marcaide J. M. and  Weiler K. W. (eds),
in press (astro-ph/0309774)

\bibitem[Kn\"odlseder (1997)]{knoedlseder97}
Kn\"odlseder J., 1997, PhD Thesis, Univ. Paul Sabatier (Toulouse),
unpublished

\bibitem[Kn\"odlseder (1999)]{knoedlseder99}
Kn\"odlseder J., 1999, \apj 510, 915

\bibitem[Kn\"odlseder (2001)]{knoedlseder01}
Kn\"odlseder J., et al. 2001, in Proceedings of the 4th INTEGRAL
Workshop, Gimenez A., Reglero V. ad Winkler C. (eds) ESA SP-459,
p. 23

\bibitem[Limongi (2003)]{limongi03}
Limongi M., Chieffi A., 2003, \apj 592, 404

\bibitem[Mahoney (1982)]{mahoney 82}
Mahoney W., Ling J.,  Jacobson A., Lingenfelter R., 1984, \apj 262, 742

\bibitem[Meynet (1997)]{meynet97}
Meynet G., Arnould M., Prantzos N., Paulus G., 1997, \aa 320, 460

\bibitem[Meynet (2003)]{meynet03}
Meynet G., Maeder A., 2003, \aa 404, 975

\bibitem[Prantzos (1991)]{prantzos91}
Prantzos N., 1991, in Gamma-ray line astrophysics, Durouchoux Ph.
and  Prantzos N. (eds), AIP, New York, p. 129

\bibitem[Prantzos (1993)]{prantzos93}
Prantzos N., 1993, \apj 405, L55

\bibitem[Prantzos (2002)]{prantzos02}
Prantzos N., 2002, in The Gamma Ray Universe, Goldwurm A., Neumann D.
and Tran Thanh Van (eds), The Gioi Publishers, Vietnam, p. 263  
(astro-ph/0209303)

\bibitem[Prantzos (1986)]{prantzos86}
Prantzos N., Cass\'e M., 1986, \apj 307, 324

\bibitem[Prantzos (1996)]{prantzos96}
Prantzos N., Diehl R., 1996, PhysRep 267, 1

\bibitem[Prantzos (1987)]{prantzos87}
Prantzos N., Arnould M., Arcoragi J. P., 1987, \apj, 315, 209

\bibitem[Rauscher (2002)]{rauscher02}
Rauscher T., Heger A., Hofmann R., Woosley S., 2002, \apj, 576,
323

\bibitem[Smith(2003a)]{smith03a}
Smith D.M., 2003a, \apj 589, L55

\bibitem[Smith(2003)]{smith03}
Smith D.M., 2003b, in Proceedings of the 3d Workshop on Astronomy with
Radioactivities, Diehl R. et al. (eds), New Astr. Rev., in press

\bibitem[Thielemann(1996)]{thielemann96}
Thielemann K.-F., Nomoto K., Hashimoto M., 1996, \apj 460, 408

\bibitem[Timmes(1995)]{timmes95}
Timmes F., Woosley S., Hartmann D., Hoffman R., Weaver T.,
Matteucci F., 1995, \apj 449, 204

\bibitem[Vuissoz (2003)]{vuissoz03}
Vuissoz C., Meynet G., Kn\"odlseder J., et al., 2003, in
Proceedings of the 3d Workshop on Astronomy with Radioactivities,
to appear in New Astr. Rev.(astro-ph0311091)

\bibitem[Woosley(2003)]{woosley03}
Woosley S., Heger A., Rauscher T., Hoffman R., 2003, 
Nuclear Physics A718, 3c

\bibitem[Woosley(1995)]{woosley95}
Woosley S., Weaver T., 1995, \apjs 101, 181

\bibitem[Woosley(1995b)]{woosley95b}
Woosley S., Langer N., Weaver T., 1995, \apj 448, 315

\end{thebibliography}

\end{document}